\documentstyle[12pt]{article}
\newlength{\dinwidth}
\newlength{\dinmargin}
\begin{document}

\def\thefootnote{\fnsymbol{footnote}}
\baselineskip18pt
\thispagestyle{empty}
\begin{flushright}
\begin{tabular}{l}
FTUAM-92/01\\\vspace*{24pt}
January, 1992
\end{tabular}
\end{flushright}

\vspace*{1.5cm}

{\vbox{\centerline{{\Large{\bf STRINGS BELOW THE PLANCK SCALE
%}}}\vskip12pt\centerline{{\Large{\bf AT}}}
%\vskip12pt\centerline{{\Large{\bf FINITE TEMPERATURE
}}}}}

\vskip72pt\centerline{M.A.R. Osorio\footnote{Bitnet address:
OSORIO@EMDUAM11} and M. A. V\'azquez-Mozo\footnote{Bitnet address:
MAVAZ@EMDUAM11}}

\vskip12pt
\centerline{{\it Departamento de F\'{\i}sica Te\'orica C-XI}}\vskip2pt
\centerline{{\it Universidad Aut\'onoma de Madrid}}\vskip2pt
\centerline{{\it 28049 Madrid, Spain}}

\vskip .7in

\baselineskip24pt
\noindent

%%%%%%%%%%%%%%%%%%%%%%%%% Abstract %%%%%%%%%%%%%%%%%%%%%%%%%%%%%%%%%%%

We show that, for a class of critical strings in ${\bf R}\times
S^{1}$-target space, the description of string theory given by its
field
content (analog model) breaks down when the radius of $S^{1}$ decreases
below $R_{0}=\sqrt{\alpha^{\prime}}$, the self-dual point of the
partition function $Z(R)$. We find that $Z(R)$ has a soft
singularity at $R_{0}$ (a finite jump in the first derivative of $Z$).

\setcounter{page}{0}
\newpage\baselineskip18pt

\section{Introduction}

The study of $c=1$ conformal matter coupled to the Liouville mode  is
equivalent to that of a bosonic string in two dimensions.
Closed bosonic string theory in two target-space dimensions can be
formulated provided we include non-trivial backgrounds
\cite{Polchinski}. The only propagating field in this string theory
is the scalar tachyon mode which actually is massless in two dimensions.

The computation of the genus-one path integral \cite{Sakai,Bershadsky}
further supports the fact that the $c=1$ theory can be described by a
massless
field in two dimensions. However a subtlety must be taken into account.
The zero mode of the free scalar field compactified on a circle produces
a one loop vacuum energy for the theory (as the result of going from
$S^1$ to $\bf{R}$ taking  $R\longrightarrow \infty$) which does not
correspond to the ultraviolet divergent vaccum energy of a quantum
field. $R$-duality actually implies that this finite vacuum energy
equals the coefficient which multiplies the $R^{-2}$ dependence in the
free energy for a single boson in two dimensions \cite{Osorio}.

This letter will be devoted to the study of a class
of critical
strings in ${\bf R}\times S^1$-target space. In this case we will
show that the description of string theory by its field content
breaks down when the radius of the compactified dimension
decreases below the self-dual point ($R=\sqrt{\alpha^{\prime}}$, we
will take $\alpha^{'}=\frac{1}{2}$ throughout). More generally,
we will see that the transition across the self-dual point physically
corresponds to the fact that, below the Planck scale, the partition
function of the string measures degrees of freedom which does not have a
quantum-field theoretical counterpart.

In the concluding Sec. 4 a mathematical relation between the vacuum
energy of the $c=1$ non-critical string and the partition function of
the states with the symmetry encoded in the Monster group $F_{1}$ will
be given.

\section{Two Dimensional Heterotic Strings on ${\bf R}\times S^{1}$}

There is a class of two-dimensional heterotic strings whose
left-moving sector can be constructed
taking the left-moving part of the bosonic string compactified on a 24
dimensional torus defined by a Niemeier lattice \cite{Enrique1}.
The
partition function for the left-moving states in the internal
torus reads
\begin{equation}
Z_{L}=\frac{\Theta_{\Gamma}}{\eta^{24}}=j(\tau)+r_{\Gamma}(1)-720
\end{equation}
where $j(\tau)$ is the modular invariant function of the modular
parameter $\tau=\tau_{1}+i\tau_{2}$  normalized in
such a way that the coefficient of $q^{-1}$  ($q=e^{2\pi i\tau}$) is
1 \cite{Koblitz} and $r_{\Gamma}(1)$, which parametrizes 24 models,
is the number of lattice vectors with (length)$^2=2$.

To get the right-moving part contribution to the partition function, we
first compactify the eight bosonic
right-moving coordinates of the supersymmetric string on the only
eight-dimensional self-dual
even lattice, $\Gamma_8$; then we break supersymmetry along the lines of
\cite{GinspargVafa}. To be more concrete, we mod out the theory by
$(-1)^F\vartheta$
where $\vartheta$ corresponds to shifting $\Gamma_{8}$ by half of a
lattice vector. We take $\vartheta=e^{2\pi i P \cdot \delta}$ with
$\delta=((\frac{1}{2})^{4},0^{4})$.
Now we have a theory in which there are
contributions from the four SO(10) conjugacy classes. The result is
\begin{equation}
Z_{R}=48+
\frac{1}{4 \overline{\eta}^{~12}}[\overline{\theta}^{~8}_{2}+
\overline{\theta}^{~8}_{3}+
\overline{\theta}^{~8}_{4}][\overline{\theta}^{~4}_{3}
-\overline{\theta}^{~4}_{4}- \overline{\theta}^{~4}_{2}]
\end{equation}
{}From this expression we see that supersymmetry is only broken  at
the massless level. Using Jacobi's ``aequatio'' the total partition
function is
\begin{equation}
Z \sim -48L^{2}\int_{\cal F}
\frac{d^2\tau}{\tau_{2}^{2}}[j(\tau)+r_{\Gamma}(1)-720]
\end{equation}
where $L \longrightarrow \infty$ regularizes the constant modes in the
uncompactified dimensions and ${\cal F}$ is the fundamental region of
the modular group
\begin{equation}
{\cal F}=\{\tau=\tau_{1}+i\tau_{2} \in {\cal H} \hspace{2mm} |
\hspace{2mm} -\frac{1}{2} \leq \tau_{1}<\frac{1}{2},
|\tau|>1\} \end{equation}
When we consider these models in a target-space of the form
${\bf R} \times S^{1}$ the generic partition function can be written
\begin{equation}
Z(R) \sim -48 L (2\pi R)\int_{\cal F}
\frac{d^2\tau}{\tau_{2}^{2}}\sum_{m,n} \exp\left\{-\frac{2\pi
R^{2}}{\tau_{2}}|m\tau+n|^{2}\right\} [j(\tau)+r_{\Gamma}(1)-720]
\label{eq:suma}
\end{equation}

We shall now study the analyticity properties of the
partition function.
It is easy to see from the Fourier series for
$j(\tau)$
\begin{equation}
j(\tau)=q^{-1}+744+\sum_{n=1}^{\infty} c_{n}q^{n} \hspace{2cm} c_n \in
{\bf Z}
\end{equation}
that any non-analytic behaviour in the partition function may
come exclusively from the term $q^{-1}$. This term would give a
divergent contribution for the zero mode of the solitonic sum if we do
not take
the prescription of performing first the $\tau_{1}$ integration in a
vicinity
of infinity (i.e. a region $U_{\epsilon}$ defined as the set of points
in ${\cal
F}$ such that $\tau_{2}>\epsilon \geq 1$). For the $R$-dependent part
we apply the same prescription. After performing the $\tau_{1}$
integration
on $U_{\epsilon}$ of the term $q^{-1}\times \sum_{m,n}$ only two terms
survive,
namely those related to $m=1$, $n=-1$ and $m=-1$, $n=1$ in the sum.
So the only possible non-analytic term in the sum will be proportional
to
\begin{equation}
2\int_{\epsilon}^{\infty}\frac{d\tau_{2}}{\tau_{2}^{3/2}}e^{-\pi
\tau_{2}\left(2R^{2}+\frac{1}{2R^{2}}-2 \right)}
\end{equation}
Since the argument of the exponential has a double root at
$R_{0}=\frac{1}{\sqrt{2}}$ we can write the last expression as
\begin{equation}
2\int_{\epsilon}^{\infty}\frac{d\tau_{2}}{\tau_{2}^{3/2}}e^{-\pi
\tau_{2}\left(\frac{R^{2}-R^{2}_{0}}{RR_{0}} \right)^{2}}
\end{equation}
In spite of its naive appearance, the analiticity properties of this
integral as a function of $R$ are nontrivial.  One can
easily see that it is finite and continuous at $R=R_{0}$, but its first
derivative with respect to $R$ is discontinuous at that point. Denoting
by $I_{\epsilon}(R)$ the last integral, we have
\begin{equation}
\frac{dI_{\epsilon}}{dR}(R) = -\frac{4\pi}{R_{0}}\left
(\frac{R^{2}-R^{2}_{0}}{RR_{0}}
\right) \left(\frac{R^{2}+R^{2}_{0}}{R^{2}} \right)
\int_{\epsilon}^{\infty} \frac{d\tau_{2}}{\tau_{2}^{1/2}}
 e^{-\pi \tau_{2} \left(\frac{R^{2} - R^{2}_{0}}{RR_{0}}\right)^{2}}
\end{equation}
Now the integral can be calculated for any $R \neq R_{0}$ by
performing a change of variables, the result being
\begin{equation}
\frac{dI_{\epsilon}}{dR}(R)=-\frac{4\sqrt{\pi}}{R_{0}}
\left(\frac{R^{2}+R_{0}^{2}}{R^{2}}\right) sig(R-R_{0})
\Gamma\left(\frac{1}{2},\pi \epsilon
\left(\frac{R^{2}-R_{0}^{2}}{RR_{0}}\right)^{2}\right)
\end{equation}
where $\Gamma (a,x)$ is the incomplete gamma function and $sig(x)$ the
sign function. The jump in the first derivative is then given by
\begin{equation}
\frac{dI_{\epsilon}}{dR}(R_{0}^{+})-\frac{dI_{\epsilon}}{dR}(R_{0}^{-})
= -\frac{16\pi}{R_{0}}
\end{equation}
Since any other term in (\ref{eq:suma})  is
analytic for any $R$, the partition function itself suffers from a
discontinuity in the first derivative at the  point  $R_{0}$.
This is the only point in which the partition function is not analytic.

Let us study now the $R$-duality properties of $Z(R)$.
It is easy to see, applying Poisson resummation formula,
that the solitonic sum is invariant under the replacement
\begin{equation}
R \longrightarrow \frac{1}{2R}
\end{equation}
so the partition function satisfies
\begin{equation}
Z(R)=Z\left(\frac{1}{2R}\right)
\end{equation}
the critical point is thus the self-dual point, $R_{0}=R_{self-dual}$.

\section{Massless Fields in ${\bf R}\times S^{1}$ and Strings}

The vacuum energy per degree of freedom of a bosonic massless field in
${\bf R}\times S^{1}$ is given by \cite{Enrique2,Osorio}
\begin{equation}
\Lambda(R)=-L(2\pi R)(2\pi)^{-2} \int_{0}^{+\infty}\frac{ds}{s^{2}}
\left.\sum_{r=-\infty}^{+\infty}\right.^{\prime} \exp{\left\{-\frac{2\pi
R^{2}}{s} r^{2}\right\}} + \Lambda_{0} = -\frac{L}{12R} + \Lambda_{0}
\label{eq:field}
\end{equation}
where $\Lambda_{0}$ is the ultraviolet divergent vacuum energy of the
field in ${\bf R}^{2}$ that would correspond to the $r=0$ term in the
sum.
As was done in \cite{Sakai,Bershadsky} we introduce the unity in
the form
\begin{equation}
1=\int_{-\frac{1}{2}}^{+\frac{1}{2}}d\theta
\end{equation}
and apply the techniques of \cite{Maclain,O'Brien,Enrique3} to get the
following relationship
\begin{equation}
\int_{0}^{+\infty}ds\int_{-\frac{1}{2}}^{+\frac{1}{2}}d\theta {s^{-2}}
\left.\sum_{r=-\infty}^{+\infty}\right.^{\prime} \exp{\left\{-\frac{2\pi
R^{2}}{s} r^{2}\right\}} = \int_{\cal F} \frac{d^{2} \tau}{\tau_{2}}
\left.\sum_{m,n\in {\bf Z}}\right.^{\prime} \exp{\left\{-\frac{2\pi
R^{2}}{\tau_{2}}{|m\tau+n|}^{2}\right\}}
\label{eq:rel}
\end{equation}
where we have defined $\tau=\tau_{1}+i\tau_{2}=:\theta+is$. This means
that the temperature dependent one-loop free energy for a single
massless boson in two dimensions (with the indentification $1/T=2\pi R$)
is equivalent to the temperature dependent Helmholtz free energy of the
continuum Liouville theory coupled to conformal matter with $c=1$.

Let us now introduce the unity as
\begin{equation}
1=\frac{1}{744}\int_{-\frac{1}{2}}^{+\frac{1}{2}}j(\tau)d\tau_{1}
\end{equation}
Then we would have the following  would-be equality
\begin{eqnarray}\lefteqn{
\tilde{\Lambda}_{1}(R)=:\int_{\cal F} \frac{d^{2} \tau}{\tau_{2}^2}
\left.\sum_{m,n\in {\bf Z}}\right.^{\prime} \exp{\left\{-\frac{2\pi
R^{2}}{\tau_{2}}{|m\tau+n|}^{2}\right\}}\stackrel{?}{=}}\nonumber\\
& &{\frac{1}{744}}
\int_{\cal F} \frac{d^{2} \tau}{\tau_{2}^2}j(\tau)
\left.\sum_{m,n\in {\bf Z}}\right.^{\prime} \exp{\left\{-\frac{2\pi
R^{2}}{\tau_{2}}{|m\tau+n|}^{2}\right\}}=:\tilde{\Lambda}_{2}(R)
\label{eq:clave}
\end{eqnarray}
If we included the zero mode in both of the double sums we would see
that  (\ref{eq:clave}) would not be correct because
\begin{equation}
\int_{\cal F} \frac{d^{2} \tau}{\tau_{2}^2}  \neq
\frac{1}{744}\int_{\cal F} \frac{d^{2} \tau}{\tau_{2}^2}j(\tau) =
\frac{720}{744}\int_{\cal F} \frac{d^{2} \tau}{\tau_{2}^2}
\label{eq:zero}
\end{equation}
$R$-duality symmetry for each of the integrals in
(\ref{eq:clave}), which is actually a property
of the solitonic sector in the integrand, reads \cite{Osorio}
\begin{equation}
\tilde{\Lambda}_{i}(R)=\frac{1}{2R^{2}}\tilde{\Lambda}_{i}(\frac{1}{2R})
-(1-\frac{1}{2R^{2}})\times
(zero\hspace{6pt}mode\hspace{6pt}contribution)_{i}
\label{eq:duality}
\end{equation}
where $\tilde{\Lambda}_{i}$ represents any of the two integrals in
(\ref{eq:clave}) and
$(zero\hspace{6pt}mode\hspace{6pt}contribution)_{i}$
is given respectively by the two sides of the inequality
(\ref{eq:zero}). Therefore $R$-duality symmetry implies that the
integrals in (\ref{eq:clave}) cannot be equal for any value of $R$.
To be more concrete, since $\lim_{R\rightarrow
0^{+}}\tilde{\Lambda}_{i}(R)$ goes as $(1/R^{2})\times
(zero\hspace{6pt}mode\hspace{6pt}contribution)_{i}$,
(\ref{eq:clave}) is an inequality for ``small'' $R$.
This implies that we cannot naively use the theorem of
\cite{Maclain,O'Brien,Enrique3}. The main subtlety in our case is the
fact that on the left hand side of
\begin{eqnarray}\lefteqn{
\frac{1}{744}\int_{0}^{+\infty}{s^{-2}}ds
\left[\int_{-\frac{1}{2}}^{+\frac{1}{2}}
j(\theta+is)d\theta\right]\left.\sum_{r=-\infty}^{+\infty}
\right.^{\prime}
\exp{\left\{-\frac{2\pi
R^{2}}{s} r^{2}\right\}} =}\\
& &\frac{1}{744}\int_{\cal F}
\frac{d^{2}\tau}{\tau_{2}^2} j(\tau)\left.\sum_{m,n\in {\bf
Z}}\right.^{\prime}
\exp{\left\{-\frac{2\pi
R^{2}}{\tau_{2}}{|m\tau+n|}^{2}\right\}}\nonumber
\label{eq:jrel}
\end{eqnarray}
a precise order of integration has been taken \cite{Kutasov}. The
application of the theorem of \cite{Maclain,O'Brien,Enrique3} to this
case must be seen as a redefinition of the integral on the planar region
$S=\{\tau | \tau_{2}>0, -\frac{1}{2}\leq \tau_1< \frac{1}{2}\}$
\cite{Osorio} in a way consisting in mapping the region $\{S-F\}$ into
$F$ an infinite number of times plus a prescription about how to
integrate over a neighborhood of infinity in each of these infinitely
many
copies of $F$. So we have an infinite number of changes of variables
(labeled by a couple of coprime integer numbers) and a prescription that
by inverting the mappings would correspond to a way of integrating in
the vicinity of zero. The validity of these changes of variables depends
on the behaviour of the integrand on the left
hand side of the equation (\ref{eq:jrel}) as a function of the modular
parameter. From the modular invariance of
$j(\tau)$ one can conclude that when $R>1$ the integrand is
regular (bounded)
at zero. Therefore this complicated way of obtaining the integral over
$S$ will give the Riemann integral which is equal to the iterated
integration at least in the region in which $R>1$.
Furthermore, the
fact that we know that $\tilde{\Lambda}_{2}(R)$ has only one singular
point means that $\tilde{\Lambda}_{1}(R)=\tilde{\Lambda}_{2}(R)$ for
$R\geq R_{0}$. Taking this together with (\ref{eq:duality}) we get that
$\tilde{\Lambda}_{1}(R)>\tilde{\Lambda}_{2}(R)$ for $R<R_{0}$.

\section{Conclusions}

We have shown that the description of string theories as collections of
fields breaks down when the typical length  of the target space is
smaller than the Planck length, $\sqrt{\alpha^{\prime}}$. In the cases
we have treated the physical explanation for this phenomenon can be
enlighted by seeing that equation (\ref{eq:clave}) is equivalent to
knowing when
\begin{equation}
M(R)=:\int_{\cal F}\frac{d^{2}\tau} {\tau_{2}^2}
J(\tau)\left.\sum_{m,n\in {\bf Z}}\right.^{\prime}
\exp{\left\{-\frac{2\pi
R^{2}}{\tau_{2}}{|m\tau+n|}^{2}\right\}}\stackrel{?}{=}0
\label{eq:Monster}
\end{equation}
where $J(\tau)=j(\tau)-744$. We have proven that this integral vanishes
when $R\geq R_{0}$ and, by duality symmetry that
\begin{equation}
M(R)=(1-\frac{1}{2R^{2}})\frac{24\pi}{3}\hspace{1cm}R\leq R_{0}
\end{equation}
Moreover, the jump of the first derivative at $R_{0}$ can be easily
seen equal to $16\pi/R_{0}$ as computed in section 2. For all of our
examples
this stems from the fact that duality in addition with the description
of the string as a collection of quantum fields gives
the partition function as a function of $R$ for all $R>0$.

Physically, $M(R)$ when $R\geq R_{0}$ would measure the quantum-field
degrees of freedom of the model resulting from taking the
left-moving part of the bosonic string after compactifying it using the
Leech lattice \cite{Enrique1} and modding out the states at the massless
level \cite{Harvey,Enrique1}. In other words, $M(R)$  measures nothing
from a quantum-field theoretical point of view. However in
\cite{Lepowsky} it has been shown that $J(\tau)$ is the
partition function corresponding to a string realization of the symmetry
encoded in the finite group $F_{1}$ usually called the Monster. What we
see is that below the Planck length strings have physical degrees of
freedom which correspond to non-propagating states.
Therefore, it appears that Atkin-Lehner symmetry
\cite{Moore,Enrique1} might be, as a stringy effect, more fundamental
than previously thought.

One may guess that the one loop vacuum energy of the continuum
Liouville theory coupled to conformal matter with $c=1$
stems from the contribution of the so-called "co-dimension two" states
\cite{Polyakov} whose propagator cannot be defined. Neither we
know how to prove this guess nor
whether there is any relationship between the kind of physical
information in $F_{1}$ and these states.
What we know is that
\begin{equation}
\int_{\cal F}\frac{d^{2}\tau}{\tau_{2}^{2}}=
 -\frac{1}{24}\int_{\cal F}\frac{d^{2}\tau}{\tau_{2}^{2}}J(\tau)
\end{equation}
namely, we have a relation between the vacuum energy corresponding to
the $c=1$ non-critical string and the Monster group.

\section*{Acknowledgements}

We would like to thank  E. \'Alvarez for a
careful
reading of the manuscript and many  useful suggestions.
We also wish to thank Jos\'e Luis F. Barb\'on for fruitful discussions
that have improved the final outcome of the present work.
M. A. V.-M. wishes to dedicate this work to the memory of
Jos\'e V\'azquez Oliver.

\end{document}